\documentclass[prl,showpacs,amsmath,amssymb, twocolumn, 10pt]{revtex4}

\usepackage{amsthm}
\usepackage{dcolumn}
\usepackage{bm}
\usepackage{graphicx}

\begin{document}

\newtheorem{corollary}{Corollary}
\newtheorem{definition}{Definition}
\newtheorem{example}{Example}
\newtheorem{lemma}{Lemma}
\newtheorem{proposition}{Proposition}
\newtheorem{theorem}{Theorem}
\newtheorem{fact}{Fact}
\newtheorem{property}{Property}
\newcommand{\bra}[1]{\langle #1|}
\newcommand{\ket}[1]{|#1\rangle}
\newcommand{\braket}[3]{\langle #1|#2|#3\rangle}
\newcommand{\ip}[2]{\langle #1|#2\rangle}
\newcommand{\op}[2]{|#1\rangle \langle #2|}

\newcommand{\tr}{{\rm tr}}
\newcommand {\E } {{\mathcal{E}}}
\newcommand {\F } {{\mathcal{F}}}
\newcommand {\diag } {{\rm diag}}

\title{Local Distinguishability of Multipartite Unitary Operations}
\author{Runyao Duan}
\email{dry@tsinghua.edu.cn}
\author{Yuan Feng}
\email{feng-y@tsinghua.edu.cn}
\author{Mingsheng Ying}
\email{yingmsh@tsinghua.edu.cn}

\affiliation{State Key Laboratory of Intelligent Technology and
Systems, Department of Computer Science and Technology, Tsinghua
University, Beijing, China, 100084}

\date{\today}

\begin{abstract}
We show that any two different unitary operations acting on an
arbitrary multipartite quantum system can be perfectly
distinguishable by local operations  and classical communication
when a finite number of runs is allowed. We then directly  extend
this result into the case when the number of unitary operations to
be discriminated is more than two. Intuitively, our result means
that the lost identity of a nonlocal (entangled) unitary operation
can be recovered locally, without any use of entanglement or joint
quantum operations.
\end{abstract}


\maketitle

Unitary operation is one of the most fundamental ingredients  of
quantum mechanics. The study of various properties of unitary
operations lies at the heart of many  quantum information processing
tasks. Recently the discrimination of unitary operations has
received many attentions \cite{CP00,AC01,DPP01,DFY07}. As a matter
of fact, the well-known effect of quantum super-dense coding
\cite{BS92} can be treated as an instance of the discrimination of
unitary operations \cite{CP00,OR04,MOR05}. Although two
nonorthogonal quantum states cannot be perfectly distinguishable
whenever only a finite number of copies are available\cite{CHE01,
ACM+07}, it was shown that any two different unitary operations, no
matter orthogonal or not, can always be perfectly distinguishable by
taking a suitable entangled state as input and then applying only a
finite number of runs of the unknown unitary operation
\cite{AC01,DPP01}. This result was further refined by showing that
the entangled input state is not necessary \cite{DFY07}. The
probabilistic discrimination of unitary operations as well as
general quantum operations has also been studied extensively
\cite{CS03,CKT+07,SAC05,WY06,JFDY06}.

\begin{figure}[ht]
  \centering
  \includegraphics[scale=0.4]{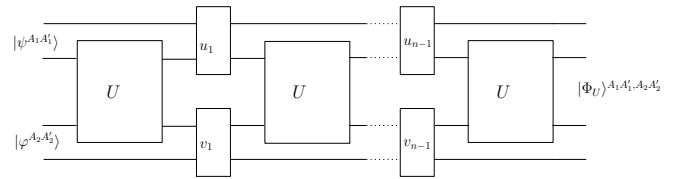}
  \caption{Illustration of LOCC discrimination of unitary operations: A bipartite example.
  Here $U\in \{U_1,U_2\}$ represents the unknown bipartite unitary
  operation. $u_k$ and $v_k$ are the local unitary operations
  performed by Alice and Bob, respectively. A general scheme for Alice and Bob to
  identify $U$ is as follows: (1). Prepare suitable input states
  $\ket{\psi}^{A_1A_1'}$ and $\ket{\varphi}^{A_2A_2'}$ as respective input states, where $A_1'$ and $A_2'$ are the auxiliary quantum systems
  of Alice and Bob, respectively; (2). Execute a finite number of runs of $U$ and insert
  appropriate local unitary operations between every two successive
  runs; (3). Distinguish the final output states
  $\ket{\Phi_U}^{A_1A_1',A_2A_2'}$ by LOCC. $U_1$ and $U_2$ can be perfectly distinguishable  if and only if
  the final output states $\ket{\Phi_{U_1}}$ and $\ket{\Phi_{U_2}}$ can be orthogonal \cite{WSHV00}.}
  \label{locc}
\end{figure}

Up to now all the above discrimination schemes of quantum operations
assume that the unknown quantum operation to be discriminated is
under the completely control of a single party who can prepare any
entangled states or perform any unconstrained quantum measurements
in order to achieve an optimal discrimination. However, any
reasonable quantum system in practice generally consists of several
subsystems. Nonlocal unitary operations are a valuable resource to
interact different subsystems together \cite{ZZF00, BC01, VHC02,
NDD+03}. The problem of distinguishing multipartite unitary
operations naturally arises when several parties share a unitary
operation but forget the real identity of the operation.
Fortunately, they do remember that the unknown unitary operation
belongs to a finite set of pre-specified unitary operations.  As in
this scenario different parties may be far from each other, a
reasonable constraint on the discrimination is that each party is
only allowed to perform local operations and classical communication
(LOCC). Moreover, we assume that there is no pre-shared entanglement
between any two distant parties. Here we may have two kinds of
entanglement: One is shared between distant parties and the other is
existing between different subsystems of a same party. The most
expensive entanglement we are concerned with is the former and the
latter can be used in order to achieve an optimal discrimination.  A
general scheme for LOCC discrimination of unitary operations is
intuitively depicted as Fig. \ref{locc}. Two special kinds of
schemes are of particular interests. A scheme is said to be {\it
parallel} if the computational network in Fig. \ref{locc} is reduced
to the form of $U^{\otimes N}$ for some finite $N$. While it is said
to be {\it sequential} if no auxiliary quantum systems are involved.
In other words, in a sequential scheme every party cannot employ
local entanglement and can only perform local unitary operations and
projective measurements on  a single quantum system. Clearly, a
sequential scheme represents the most economic strategy for
discrimination.

The purpose of this Letter is to show that any two multipartite
unitary operations can be perfectly distinguishable even under the
constraint of LOCC. Our scheme for discrimination is rather simple
as it only involves with parallel scheme and sequential scheme and
only requires  one party to prepare local entanglement. By similar
arguments as that in Refs. \cite{AC01,DFY07}, we can directly extend
this result into the case when the number of the unitary operations
to be discriminated is more than two.  It is remarkable that the
lost identity of a nonlocal unitary operation can be recovered
locally without the assistance of any {\it a priori} entanglement.
To our knowledge, this is the first result about the local
distinguishability of multipartite quantum operations.  An immediate
application is as follows. Suppose several parties share an unknown
unitary operation which is secretly chosen from a finite set of
unitary operations, each of which is assumed to be capable of
creating entanglement locally. Then these parties can always produce
pure multipartite entanglement with certainty by employing the
unknown operation shared among them. On the other hand, the same
task is not possible if we consider the distillation of
nonorthogonal entangled states instead of unitary operations.

Obviously, the proof presented in this Letter automatically provides
an alternative way to show the perfect distinguishability between
unitary operations in the global scenario \cite{AC01,DPP01,DFY07}.
However, due to the nonlocal nature of general multipartite unitary
operations, the proof for the local distinguishability is rather
complicated and needs lots of new techniques. For instance, the
notion of numerical range for a linear operation has been
generalized to multipartite setting and many interesting properties
are presented. We hope these tools would also be useful in studying
other problems in quantum information theory.

Let us begin to introduce the notion of numerical range. Consider a
quantum system associated with a finite dimensional state space
$\mathcal{H}$. The set of linear operations acting on $\mathcal{H}$
is denoted by $\mathcal{B}(\mathcal{H})$. In particular,
$\mathcal{U}(\mathcal{H})$ is the set of unitary operations acting
on $\mathcal{H}$. Two unitary operations $U,V\in
\mathcal{U}(\mathcal{H})$ are said to be different if
$U=e^{i\theta}V$ cannot hold for any real number $\theta$. For $A\in
\mathcal{B}(\mathcal{H})$. The numerical range (or the field of
values) of $A$ is a subset of complex numbers defined as follows:
\begin{equation}\label{WA}
W(A)=\{\braket{\psi}{A}{\psi}:\ip{\psi}{\psi}=1\}.
\end{equation}

When $A$ is a normal operation, i.e., $AA^\dagger=A^\dagger A$. By
spectral decomposition theorem it is easy to verify that
$W(A)=Co(\sigma(A))$, where $\sigma(A)$ represents the set of
eigenvalues of $A$ and $Co(S)$ denotes the convex hull of $S$ for
$S\subseteq \mathcal{C}$. In other words, the numerical range of a
normal operation is a convex polygon. Unfortunately, no similar
analytical characterization of numerical range is known for general
linear operations. Nevertheless, a celebrated theorem due to
Toeplitz and Hausdorff states that the numerical range of a bounded
linear operator is always convex. For our purpose here, a finite
dimensional version of this theorem is sufficient \cite{HJ1}.
\begin{lemma}\label{th}\upshape
For any $A\in \mathcal{B}(\mathcal{H})$, $W(A)$ is convex. Moreover,
let $\{\ket{\psi_k}\}$ be a finite set of normalized states, and let
$\{p_k\}$ be a probability distribution, then the state $\ket{\psi}$
such that $\braket{\psi}{A}{\psi}=\sum_{k}p_k
\braket{\psi_k}{A}{\psi_k}$ can be chosen as a linear combination of
$\ket{\psi_k}$, i.e, $\ket{\psi}\in {\rm span}\{\ket{\psi_k}\}$.
\end{lemma}

If $\ket{\psi}$ in Eq. (\ref{WA}) can be made entangled, then we can
define the entanglement-assisted numerical range of $A$ as follows:
\begin{equation}\label{AWA}
W_a(A)=\cup_{\mathcal{H}'} W(A\otimes I_\mathcal{H'}),
\end{equation}
where $\mathcal{H}'$ ranges over all finite dimensional state
spaces. One can  verify by a direct calculation that
$$W_a(A)=\{\tr(A\rho):\rho\geq 0,
\tr(\rho)=1\}.$$ It follows from Lemma \ref{th} that
$W_a(A)=Co(W(A))=W(A)$ for any $A\in\mathcal{B}(\mathcal{H})$.

Suppose now we are concerned with a multipartite quantum system
consisting of $m$ parties, say, $M=\{A_1,\cdots,A_m\}$. Assume that
the party $A_k$ has a state space $\mathcal{H}_k$ with dimension
$d_k$. Then the whole state space is given by
$\mathcal{H}=\otimes_{k=1}^m\mathcal{H}_k$ with total dimension
$d=d_1\cdots d_m$. We often use $d_1\otimes\cdots\otimes d_m$ as an
abbreviation for $\mathcal{H}$. $U\in\mathcal{U}(\mathcal{H})$ is
said to be local or decomposable if $U=\otimes_{k=1}^m u_k$ such
that $u_k\in \mathcal{U}(\mathcal{H}_k)$. Otherwise $U$ is nonlocal
or entangled. The local numerical range of $A$ is a subset of $W(A)$
with the additional requirement that $\ket{\psi}$ in Eq. (\ref{WA})
is a product state. That is,
\begin{equation}\label{localWA}
W^{local}(A)=\{\braket{\psi}{A}{\psi}:
\ket{\psi}=\otimes_{k=1}^m\ket{\psi_k}\},
\end{equation}
where $\ket{\psi_k}\in \mathcal{H}_k$ and $\ip{\psi_k}{\psi_k}=1$.
The local entanglement-assisted numerical range $W_a^{local}(A)$ can
be defined similar to $W_a(A)$. A simple observation is as follows:
$$W^{local}_a(A)=\{\tr(A\rho):\rho=\otimes_{k=1}^m\rho_k\},$$
where $\rho_k$ is a density operator on $\mathcal{H}_k$. A rather
surprising result is that local entanglement cannot broaden the
local numerical range even in the multipartite scenario.
\begin{lemma}\label{localWAAWA}\upshape
For any $A\in \mathcal{H}$, $W_a^{loca}(A)=W^{local}(A)$.
\end{lemma}

{\bf Proof.} The proof is  a simple application of Lemma \ref{th}.
For simplicity, we only consider bipartite case. Denote
$f(\psi_1,\psi_2)=\tr(A\op{\psi_1}{\psi_1}\otimes
\op{\psi_2}{\psi_2})$. First we observe that
$f(\psi_1,\psi_2)=\braket{\psi_1}{A_{\psi_{2}}}{\psi_1},$ where
$A_{\psi_2}=\tr_{\mathcal{H}_{2}}(A I_{\mathcal{H}_1}\otimes
\op{\psi_{2}}{\psi_{2}})$. So it follows from Lemma \ref{th} and the
symmetry that $f(\psi_1,\psi_2)$ is convex in $\op{\psi_1}{\psi_1}$
(or $\op{\psi_2}{\psi_2}$) when $\op{\psi_2}{\psi_2}$ (resp.
$\op{\psi_1}{\psi_1}$) is fixed. Hence for any density operators
$\rho_1$ and $\rho_2$ there should exist pure states $\ket{\psi_1}$
and $\ket{\psi_2}$ such that $\tr(A\rho_1\otimes
\rho_2)=f(\psi_1,\psi_2)$. \hfill $\square$

We shall employ a fundamental result by Walgate {\it et al}
\cite{WSHV00} to study the local distinguishability of nonlocal
unitary operations.
\begin{lemma}\label{walgatetheorem}\upshape
({Walgate {\it et al}, \cite{WSHV00}): }Let $\ket{\psi_1}$ and
$\ket{\psi_2}$ be two multipartite orthogonal pure state on
$\mathcal{H}$. Then $\ket{\psi_1}$ and $\ket{\psi_2}$ are perfectly
distinguishable by LOCC.
\end{lemma}

The relation between local distinguishability of unitary operations
and local numerical range now is clear. Actually, if $0\in
W^{local}(U^\dagger_2 U_1)$ then there exists a product state
$\ket{\psi}$ such that $U_1\ket{\psi}$ and $U_2\ket{\psi}$ are
orthogonal. It follows from the above lemma that $U_1$ and $U_2$ can
be perfectly distinguishable by LOCC. Conversely, suppose that $U_1$
and $U_2$ can be discriminated by LOCC, then there exists a product
state $\ket{\psi}^{MM'}=\otimes_{k=1}^m \ket{\psi_k}^{A_kA_k'}$ such
that $(U_1^M\otimes I^{M'})\ket{\psi}^{MM'}$ and $(U_2^M\otimes
I^{M'})\ket{\psi}^{MM'}$ are orthogonal, where $A_k'$ is a local
auxiliary system of $A_k$. That is equivalent to $0\in
W^{local}_a(U_1^\dagger U_2)$. By Lemma \ref{localWAAWA}, this is
also equivalent to $0\in W^{local}(U_1^\dagger U_2)$. Interestingly,
local entanglement is not necessary for the perfect local
discrimination between two unitary operations.
\begin{theorem}\label{localunitary}\upshape
Two unitary operations $U_1$ and $U_2$ are perfectly distinguishable
by LOCC in the single-run scenario if and only if $0\in
W^{local}(U^\dagger_1 U_2)$.
\end{theorem}
For simplicity a state $\ket{\psi}$ such that
$\braket{\psi}{A}{\psi}=0$ is said to be an {\it isotropic vector}
for $A$. The term {\it isotropic product vector} is used when
$\ket{\psi}$ is a product state. As a simple application of Lemma
\ref{localWAAWA}, we have $\tr(U_1^\dagger U_2)=0$ implies that
$U_1^\dagger U_2$ has an isotropic product state. Hence $U_1$ and
$U_2$ are perfectly distinguishable by LOCC with a single run.

Unfortunately, how to determine when $0$ is in the local numerical
range remains unknown even for unitary operations. Consequently, it
is generally difficult to decide the local distinguishability of
nonlocal unitary operations in the single-run scenario. Since the
set of LOCC operations is very restricted, it is not clear whether
nonlocal unitary operations remain locally distinguishable. Indeed,
the following example demonstrates that the LOCC discrimination and
the global discrimination of unitary operations are very different
when only the single-run scenario is considered.

\begin{example}\label{productbasis}\upshape
Let $U_1$ and $U_2$ be $2\otimes 2$ unitary operations such that
$U_1^\dagger
U_2=\op{00}{00}+e^{i\theta_1}\op{01}{01}+e^{i\theta_2}\op{10}{10}-\op{11}{11}$
for $0<\theta_1, \theta_2<\pi$.

On the one hand, by taking
$\ket{\psi}=(\ket{00}+\ket{11})/\sqrt{2}$, we have
$\braket{\psi}{U^\dagger_1 U_2}{\psi}=0$. That implies $U_1$ and
$U_2$ are perfectly distinguishable by employing a maximally
entangled state as input. On the other hand, we can easily verify
that $U_1^\dagger U_2$ cannot have an isotropic product state, thus
$U_1$ and $U_2$ are locally indistinguishable.
\hfill $\square$\\
\end{example}

The above example also demonstrates that the local numerical range
is not convex in general. More precisely, we have $\pm 1\in
W^{local}(U^\dagger_1 U_2)$ as one can choose $\ket{\psi}$ as
$\ket{00}$ and $\ket{11}$, respectively. However, $0=(-1+1)/2\not
\in W^{local}(U^\dagger_1 U_2)$.  An interesting question is to ask
for what kind of linear operations the local numerical range remains
convex. The general answer to this question is unknown. Here we
would like to point out that such a convex property does hold for
Hermitian operations, for which the local numerical range is just a
complex segment.

Remarkably, if we are allowed to use the unknown multipartite
unitary repeatedly, then any two different multipartite unitary
operations become locally distinguishable. In what follows we shall
present a  complete proof of this interesting fact. For the ease of
presentation, the lengthy proof is divided into two parts: Theorem
\ref{mlocc} and Theorem \ref{slocc}.

Some technical lemmas are necessary in order to present such a
proof. The following useful lemma provides an alternative
characterization of Hermitian operations.
\begin{lemma}\label{Hermitian}\upshape
Let $\{\rho_k:1\leq k\leq d^2\}$ be a Hermitian basis for
$\mathcal{B}(\mathcal{H})$. Then $A\in \mathcal{B}(\mathcal{H})$ is
Hermitian if and only if ${\tr}(A\rho_k)\in \mathcal{R}$ for all
$1\leq k\leq d^2$.
\end{lemma}

There are many ways to choose a Hermitian basis. Here is a simple
construction based on the idea of quantum process tomography
\cite{CN97}. Let $\{\ket{k}:1\leq k\leq d\}$ be an orthonormal basis
for $\mathcal{H}$. For $1\leq p<q\leq d$, let
$\ket{\psi_{pq}^+}=(\ket{p}+\ket{q})/{\sqrt{2}}$ and
$\ket{\psi_{pq}^-}=(\ket{p}+i\ket{q})/{\sqrt{2}}.$ In addition, for
$1\leq p\leq d$ let $\ket{\psi_{pp}}=\ket{p}$. Then
$$\{\op{\psi_{pq}^{\pm}}{\psi_{pq}^{\pm}}:1\leq p<q\leq d\}\cup
\{\op{\psi_{pp}}{\psi_{pp}}:1\leq p\leq d\}$$ is a Hermitian basis
for $\mathcal{B}(\mathcal{H})$.

For a set of complex numbers $\{z_k\}$, $z_k$s are co-linear if
there exists $0\leq \theta<2\pi$ such that $z_k=r_ke^{i\theta}$ and
$r_k\geq 0$ for any $k$. Geometrically, $z_k$s are co-linear if they
lie on the same ray from the origin. The following lemma is crucial
in proving our main result.  Note that $\lceil x\rceil$ represents
the minimum of the integers that are not less than $x$.

\begin{lemma}\label{keylemma}\upshape
For $A\in \mathcal{B}(\mathcal{H})$, let $\ket{\psi_1}$ and
$\ket{\psi_2}$ be two normalized vectors  such that
$\braket{\psi_1}{A}{\psi_1}=r_1e^{i\theta_1}$ and
$\braket{\psi_2}{A}{\psi_2}=r_2e^{i\theta_2}$ are not co-linear,
where $r_1,r_2>0$ and $0\leq \theta_1<\theta_2<2\pi$. Define
$\theta=\min\{\theta_2-\theta_1, 2\pi+\theta_1-\theta_2\}$ and
$N=\lceil\frac{\pi}{\theta}\rceil$. Then $0\in W(A^{\otimes N})$,
and the isotropic vector $\ket{\psi}$ can be chosen from ${\rm
span}\{\ket{\psi_1}^{\otimes N-k}\ket{\psi_2}^{\otimes k}:0\leq
k\leq N\}$.
\end{lemma}
\textbf{Proof.} It is clear that $0<\theta\leq \pi$.  To be
specific, let us assume $\theta_1=0$ and $\theta_2\leq \pi$. Then
$\theta=\theta_2$. We deal with the following two cases separately:

Case 1: $\theta=\pi$. In this case we have $N=1$. Choose $0\leq
p\leq 1$ such that $pr_1-(1-p)r_2=0.$ Then we have
$p\braket{\psi_1}{A}{\psi_1}+(1-p)\braket{\psi_2}{A}{\psi_2}=0.$ By
Lemma \ref{th}, $0\in W(A)$.

Case 2: $0<\theta<\pi$. It is obvious that $N\theta<2\pi$. Define
$\ket{\Phi_k}=\ket{\psi_1}^{\otimes N-k}\ket{\psi_2}^{\otimes k},
0\leq k\leq N$ and $z_k=\braket{\Phi_k}{A^{\otimes N}}{\Phi_k}$. We
shall show that $0\in Co\{z_k:0\leq k\leq N\}$. A routine
calculation shows that
\begin{equation}\label{tensor}
z_k=r_1^{N-k}r_2^{k}e^{ik\theta}.
\end{equation}
To complete the proof in this case, it suffices to consider the
following two subcases:

Case 2a: $N\theta=\pi$. Then we have $e^{iN\theta}=-1$. Similar to
Case 1, we can choose $0\leq p\leq 1$ such that
$pr_1^{N}-(1-p)r_2^{N}=0,$ which immediately follows that
$pz_0+(1-p)z_N=0.$ By Lemma \ref{th}, $0\in W(A^{\otimes {N}})$.

Case 2b. $\pi<N\theta<2\pi$. By the assumption on $N$, we should
have $N\geq 2$ and $N\theta-\pi<(N-1)\theta<\pi$. These conditions
imply that for any positive real numbers $s_1, s_2, s_3$ we have
$0\in Co\{s_1, s_2e^{i(N-1)\theta},s_3e^{iN\theta}\}$. By Eq.
(\ref{tensor}), there exists $p_1$, $p_2$, $p_3$ such that
$$p_1z_0+p_2z_{N-1}+p_3z_{N}=0,$$ where $\sum_{k=1}^3p_k=1$ and
$p_k\geq 0$. Again, by Lemma \ref{th}, we have $0\in W(A^{\otimes
N})$.

In all the above cases, by the second part of Lemma \ref{th}, the
state $\ket{\psi}$ such that $\braket{\psi}{A^{\otimes N}}{\psi}=0$
can be chosen as a linear combination of $\ket{\Phi_k}$.
\hfill $\square$\\

With Lemma \ref{keylemma} in hand, we can show in the following
theorem that perfect discrimination between two multipartite unitary
operations $U_1$ and $U_2$ by a parallel scheme is always possible
except for a special case.
\begin{theorem}\label{mlocc}\upshape
Let $U_1$ and $U_2$ be two multipartite unitary operations such that
$U_1^\dagger U_2$ is non-Hermitian (up to some phase factor). Then
there exists a finite $N$ such that $0\in W^{local}(({U_1^\dagger
U_2)}^{\otimes N})$. That is, $U_1^{\otimes N}$ and $U_2^{\otimes
N}$ are perfectly distinguishable using LOCC.
\end{theorem}

\textbf{Proof.} We only need to seek a finite $N$ and a product
state $\ket{\psi}$ such that $\braket{\psi}{(U_1^\dagger
U_2)^{\otimes N}}{\psi}=0$. To simplify the notations, we consider
only the case when $U_1$ and $U_2$ both are bipartite unitary
operations acting on $\mathcal{H}_1\otimes \mathcal{H}_2$. The
general case can be proved similarly.  Let $\{\op{\psi_k}{\psi_k}\}$
and $\{\op{\varphi_l}{\varphi_l}\}$ be Hermitian basis for
$\mathcal{B}(\mathcal{H}_1)$ and $\mathcal{B}(\mathcal{H}_2)$,
respectively. Then $\{\op{\psi_k\varphi_l}{\psi_k\varphi_l}$ is a
Hermitian basis for $\mathcal{B}(\mathcal{H}_1\otimes
\mathcal{H}_2)$, where $1\leq k\leq d_1^2$ and $1\leq l\leq d_2^2$.

Consider $d_1^2d_2^2$ complex numbers
$$z_{kl}=\braket{\psi_k\varphi_l}{U_1^\dagger
U_2}{\psi_k\varphi_l}.$$ If all $z_{kl}$ are co-linear, then
$z_{kl}=r_{kl}e^{i\theta}$ for some $\theta\in \mathcal{R}$ and
$r_{kl}\in \mathcal{R}$. Thus all $e^{-i\theta}z_{kl}$s are real. By
Lemma \ref{Hermitian}, $e^{-i\theta}U^\dagger V$ is Hermitian. That
contradicts our assumption. So  there should exist $(k,l)\neq (p,q)$
such that $z_{kl}$ and $z_{pq}$ are not co-linear. More precisely,
let $z_{kl}=r_{kl}e^{i\theta_{kl}}$ and
$z_{pq}=r_{pq}e^{i\theta_{pq}}$, where $r_{kl}, r_{pq}>0$ and $0\leq
\theta_{kl},\theta_{pq}<2\pi$. We should have $\theta_{kl}\neq
\theta_{pq}$. Consider the value of $z_{kq}$. If $z_{kq}=0$ then we
can choose $\ket{\psi}=\ket{\psi_k\varphi_l}$ and the proof is
finished. Otherwise, write $z_{kq}=r_{kq}e^{i\theta_{kq}}$, where
$r_{kq}>0$ and $0\leq \theta_{kq}<2\pi$. Since $\theta_{kl}\neq
\theta_{pq}$, we should have either $\theta_{kq}\neq \theta_{kl}$ or
$\theta_{kq}\neq \theta_{pq}$. Without loss of generality, let us
assume $\theta_{kq}\neq \theta_{pq}$.  By Lemma \ref{keylemma},
there exists a finite $N$ such that $0\in W((U_1^\dagger
U_2)^{\otimes N})$. And the isotropic state $\ket{\psi}$ can be
chosen as a linear combination of the states
 $$\ket{\Phi_n}=\ket{\psi_k\varphi_q}^{\otimes N-n}\ket{\psi_p\varphi_q}^{\otimes n}, {\rm \ }0\leq n\leq N.$$
A key observation here  is that any vector from
$span\{\ket{\Phi_n}:0\leq n\leq N\}$ is of the form
$\ket{\psi'}\otimes\ket{\varphi_q}^{\otimes N}$, where
$\ket{\psi'}\in \mathcal{H}_1^{\otimes N}$ and
$\ket{\varphi_q}^{\otimes N}\in \mathcal{H}_2^{\otimes N}$. That is,
$\ket{\psi}$ can be taken as a product state.
\hfill $\square$\\

It is worth noting that in the above proof only one party is
required to prepare local entanglement.

However, the local discrimination between $U_1$ and $U_2$ such that
$U_1^\dagger U_2$ is Hermitian has not been involved yet. Noticing
that $U_1^\dagger U_2$ is Hermitian, we may write $U^\dagger
V=I-2P$, where $P$ is a projector satisfying $\tr(P)<\tr(I)/2$. The
only left case for $2\otimes 2$ is that $U_1^\dagger
U_2=I_{\mathcal{H}}-2\op{\Phi}{\Phi}$ for some state $\ket{\Phi}\in
\mathcal{H}$. Assume
$\ket{\Phi}=\sqrt{\lambda}\ket{00}+\sqrt{1-\lambda}\ket{11}$ for
some $1/2\leq \lambda\leq 1$. Then we have $\braket{00}{U_1^\dagger
U_2}{00}=1-2\lambda\leq 0$. On the other hand, we have that
$\tr(U^\dagger V)=2>0$. By the convexity of $W^{local}(U_1^\dagger
U_2)$, we have $0\in W^{local}(U_1^\dagger U_2)$. Combining this
with Theorem \ref{mlocc} we obtain the following interesting result:
\begin{corollary}\label{2otimes2}\upshape
Let $U_1$ and $U_2$ be two different $2\otimes 2$ unitary
operations. Then there exists a finite $N$ such that $0\in
W^{local}((U_1^\dagger U_2)^{\otimes N})$.
\end{corollary}
In other words, any two $2\otimes 2$ unitary operations can be
locally distinguishable by a parallel scheme.

In general, we can transform the case when $U_1^\dagger U_2$ is
Hermitian to the non-Hermitian case by applying a sequential scheme.
The following Lemma would be helpful in doing this transformation.

\begin{lemma}\upshape\label{projector1}
Let $A$ and $B$ be two Hermitian operations acting on $\mathcal{H}$
such that $u^{\dagger}AuB$ is Hermitian for any local unitary $u$.
Then $\tr(u^{\dagger}AuB)={\tr(A)\tr(B)}/{d}$ for any local unitary
$u$, where $d$ is the dimension of $\mathcal{H}$.
\end{lemma}

\textbf{Proof.} Let $f$ be a function defined on the set of local
unitary operations such that $f(u)=\tr(u^\dagger AuB)$. Then for
Hermitian operations $A$ and $B$, $f(u)\in \mathcal{R}$. By
continuity, the set of $f(u)$ is a real line segment or a singleton.
On the other hand, $u^\dagger AuB$ is Hermitian implies  that $A$
and $uBu^\dagger$ are simultaneously diagonalizable under some
unitary operation. Thus $f(u)=\tr(u^{\dagger}AuB)$ should be of the
form $\sum_{k=1}^{d}\lambda_{\xi(k)}\mu_k$ for some permutation
$\xi$, where $\lambda_k$ and $\mu_k$ are eigenvalues of $A$ and $B$,
respectively. Thus $f(u)$ can take at most $d!$ possible values and
should be a constant $C$ for any local unitary $u$.  To calculate
$C$ explicitly, let us choose a set of local unitary operations
$\{u_k:k=1,\cdots,d^2\}$ on $\mathcal{H}$ such that the following
identity holds:
\begin{equation}\label{depolar channel}
{1}/{d^2}\sum_{k=1}^{d^2}u_k^{\dagger}Au_k=\tr(A){I_{\mathcal{H}}}/{d},
\end{equation}
where $A$ is an arbitrary linear operation on $\mathcal{H}$.
Intuitively, Eq. (\ref{depolar channel}) represents the completely
depolarizing channel on $\mathcal{B}(\mathcal{H})$. Such local
unitary operations do exist. For instance, one may choose $\{u_k\}$
as the tensor products of the generalized Pauli matrices acting on
$\mathcal{H}_l$. It follows that
\begin{equation}
{1}/{d^2}\sum_{k=1}^{d^2}u_k^{\dagger}Au_kB=\tr(A){B}/{d}.
\end{equation}
Taking trace and noticing that $\tr(u_k{^\dagger}Au_k B)=C$ for any
$1\leq k\leq d^2$, we have $C={\tr(A)\tr(B)}/{d}$.  With that we
complete the proof of Lemma \ref{projector1}.\hfill $\square$

The following theorem deals with the case when $U_1^\dagger U_2$ is
Hermitian.
\begin{theorem}\label{slocc}\upshape
Let $U_1$ and $U_2$ be two different unitary operations acting on
$\mathcal{H}$ such that $U^\dagger_1 U_2$ is Hermitian (up to some
phase factor). Then there exists a finite $n>1$ and a sequence of
local unitary operations $u^{(1)},\cdots, u^{(n-1)}$ such that
$W_1^\dagger W_2$ is non-Hermitian, where $W_1=U_1u^{(1)}\cdots
u^{(n-1)}U_1$ and $W_2=U_2u^{(1)}\cdots u^{(n-1)}U_2$.
\end{theorem}

\textbf{Proof.} Without any loss of generality, we may assume that
$U^\dagger_1 U_2=D$ for some Hermitian $D$. It is worth noting that
$D=I-2P$ for some projector $P$. Hence we can assume that $\tr(D)$
is a positive integer strictly less than $d$. By contradiction,
suppose that for any $n>1$ and any local unitary operations
$u^{(1)},\cdots, u^{(n-1)}$, we have that $W_1^{\dagger}W_2$ is
Hermitian. Let $D^{(n)}=(U_1^{n})^\dagger U_2^n$. We shall prove
that
\begin{equation}\label{dimension}
\tr(D^{(n)})=({\tr(D)}/{d})^{n-1}\tr(D), ~n\geq 1.
\end{equation}
The case of $n=1$ holds trivially. Assume $n>1$. By the assumption
we have
\begin{equation}
(U^{n-1}uU_1)^\dagger (U_2^{n-1}uU_2)=U_1^\dagger[u^\dagger
D^{(n-1)}uU_1DU_1^\dagger]U_1
\end{equation}
 is Hermitian for any local unitary $u$. Applying Lemma
\ref{projector1} and setting $u=I_{\mathcal{H}}$ we have
$$\tr(D^{(n)})={\tr(D^{(n-1)})\tr(U_1DU_1^\dagger)}/{d}.$$
More explicitly,
$$\tr(D^{(n)})=({\tr(D)}/{d})\tr(D^{(n-1)}),~\tr(D^{(1)})=\tr(D).$$
Solving this relation we complete the proof of Eq.
(\ref{dimension}).

However, Eq. (\ref{dimension}) cannot  be true for all $n>1$. More
precisely, since $\tr(D)<d$, it is obvious that $\tr(D^{(n)})$ is a
strictly decreasing sequence with respect to $n$. Therefore for some
suitable $n$ we should have $0<\tr(D^{(n)})<1$, which contradicts
the fact that $\tr(D^{(n)})$ is a positive integer.  \hfill $\square$\\

In summary, we consider the discrimination between multipartite
unitary operations by  local quantum operations and classical
communications only, and show that a perfect discrimination in this
scenario is always possible. There are numerous open problems. For
example, it remains unsolved whether a perfect discrimination can be
achieved by merely a parallel scheme or a sequential scheme. Another
challenging problem is to determine the minimal number of the runs
needed for a perfect discrimination between two multipartite unitary
operations in the LOCC scenario. Similar problems have been
completely solved in the global scenario \cite{AC01, DPP01, DFY07}.

We thank Z.-F. Ji, G.-M. Wang,  J.-X. Chen, Z.-H. Wei, and C. Zhang
for helpful conversations. This work was partly supported by the
Natural Science Foundation of China (Grant Nos. 60621062 and
60503001) and the Hi-Tech Research and Development Program of China
(863 project) (Grant No. 2006AA01Z102).

\end{document}